\documentclass[preprint,authoryear,12pt]{elsarticle}

\makeatletter

\usepackage{amssymb}
\usepackage{graphicx}
\usepackage{amsmath}
\usepackage[labelformat=simple]{subfig} 
\usepackage{xcolor}
\usepackage{booktabs} 
\usepackage{multirow}
\usepackage{soul}
\usepackage{microtype}
\usepackage{slashbox}
\usepackage{pifont}
\usepackage{natbib}

\usepackage{lineno}

\makeatletter
\def\ps@pprintTitle{%
   \let\@oddhead\@empty
   \let\@evenhead\@empty
   \def\@oddfoot{\reset@font\hfil\thepage\hfil}   
\let\@evenfoot\@oddfoot
}
\makeatother
\journal{Expert Systems with Applications}

\begin{document}

\begin{frontmatter}



\title{Voting in Transfer Learning System for Ground-Based Cloud Classification}


\author{Mario Manzo}
\address{IT Services, University of Naples “L’Orientale”\\
              80121, Naples, Italy, mmanzo@unior.it}
\author{Simone Pellino}
\address{Department of Applied Science,\\ I.S. Mattei Aversa 81031,  M.I.U.R. Rome, Italy; simonepellino@gmail.com}



\begin{abstract}
Clouds classification is a great challenge in meteorological research. The different types of clouds, currently known and present in our skies, can produce radioactive effects that impact on the variation of atmospheric conditions, with the consequent strong dominance over the earth's climate and weather. Therefore, identifying their main visual features becomes a crucial aspect. In this paper, the goal is to adopt a pretrained deep neural networks based architecture for clouds image description, and subsequently, classification. The approach is pyramidal. Proceeding from the bottom up, it partially extracts previous knowledge of deep neural networks related to original task and transfers it to the new task. The updated knowledge is integrated in a voting context to provide a classification prediction. The framework trains the neural models on unbalanced sets, a condition that makes the task even more complex, and combines the provided predictions through statistical measures. Experimental phase on different cloud image datasets is performed and results achieved show the effectiveness of the proposed approach with respect to state of the art competitors.\end{abstract}




\end{frontmatter}


\section{Introduction}
\label{intro}
Clouds are a constant presence in our skies, combined with the role assumed by ecosystems, are also important in determining atmospheric conditions, hours of sunshine and temperature. Nowadays, dynamics atmospheric conditions, attributed to climate change, have led to an increase in attention to behavior of clouds \cite{dud}. Climate models allow to predict the climate changes, but their precision degree is currently insufficient and attributable to the alteration in the conditions determined by different phenomena. Consequently, clouds behavior prediction is important for estimating climate change. Furthermore, cloud changes are also a reason for influences on the earth radiation budget and energy balance \cite{Isccp,Radiative,CloudFeedbacks}. A great deal of effort has been made by the scientific community to acquire many datasets, but not fully managed due to a lack of devices with adequate processing resources. In recent years, however, the state of the art has seen the birth of numerous studies and analyzes that can automate clouds attributes and classes detection scientifically relevant. Due to the amount of ground cloud images, the recognition phase has been extensively studied lately in the literature. Most of the standard algorithms adopt hand-made features, such as brightness, texture, shape, and color, to represent image content but without obtaining a model generalization due to the complex distribution of data. Indeed, the visual information contained in the image are unable to accurately describe the clouds due to the large variations in appearance. Finally, the non-visual features, also known as multimodal information, obtainable from the clouds process formation such as temperature, humidity, pressure and wind speed can be of help. According to recent studies, deep learning has proven effective for image management, analysis, representation and classification also in cloud recognition field. In particular, the success of deep neural networks, applied to the image classification task, concern several interesting aspects mainly connected to the software development and the large amount of data available. Specifically, for cloud images analysis, deep neural networks are adopted both in the segmentation and detection phases. However, image content and poor data balance among classes have a decisive impact on performance, causing uncertainty in the model generalization. With purpose to address the above problems, we present a framework based on deep transfer and voting learning for cloud image classification. It is built based on three steps. A first, which performs image preprocessing operations such as resizing, essential for neural networks training. A second, which modifies and retrains multiple deep neural networks, exploiting previous knowledge. A third, which looks at the different predictions provided by the deep neural networks and combines them in order to provide the best decision in the classification phase. The main contributions about proposed framework can be summarized in some keypoints. First, a framework based on deep and voting learning designed to address the imbalance between classes in the cloud recognition task. Second, a framework built on multiple classification models based on deep transfer learning. Third, the demonstration that several models, suitably combined, can strengthen the decision in classification with respect to a single one. Finally, experimental demonstrations compared to established existing methods on the datasets recognized by field experts. The paper is structured as follows. Section \ref{relatedwork} provides an overview of state of art about clouds classification approaches. Section \ref{MM} describes in detail proposed framework. Section \ref{discussion} provides a wide experimental phase, while section \ref{conc} concludes the paper.

\section{Related Work}
\label{relatedwork}
In this section, we briefly describe the most important studies about clouds images classification in literature. In this field, numerous works address the task according to different aspects such as image characterization, segmentation algorithm application to get new descriptors, complex mechanisms of learning and classification and much more.

In \cite{MultimodalGroundBased} authors present a layer named joint fusion (JFCNN) to jointly learn two kinds of cloud features under one framework. After training the proposed JFCNN, they extract the visual and multimodal features from two subnetworks, which are based on the well known Resnet50 \cite{he2016deep} and integrate them using a weighted strategy. The architecture consists of five parts: two subnetworks, one joint fusion layer, one FC layer and the loss function. The subnetworks are used for learning cloud visual features. Authors work with two kind of extracted features, combined in multimodal way, which contain some complementary information and different characteristics of the ground-based cloud.  
 
An approach named multi-evidence and multi-modal fusion network (MMFN) is proposed in \cite{liu2020multi}. The idea is to learn extended cloud information by fusing heterogeneous features (global and local) in a unified framework. MMFN takes advantage of multiple pieces of evidence using a main network and an attentive network. In the attentive network, local visual features are extracted from attentive maps which are obtained by refining salient patterns from convolutional activation maps. Meanwhile, the main network learns multi-modal features for ground-based cloud. In order to combine the multi-modal and multi-evidence visual features, authors design two fusion layers in MMFN to incorporate multi-modal features with global and local visual features, respectively.

In \cite{DCAFS} authors propose to
use deep convolutional activations-based features (DCAFs). Cloud images are directly fed into a CNN model. Then, the features from different convolutional and FC layers are extracted through different pooling strategies. Finally, a multilabel linear support vector machine (SVM) model is used for the classification step.

A convolutional neural network model, called CloudNet, for accurate ground-based meteorological cloud classification is proposed in \cite{CloudNet}. The model consists in five convolutional layers and two FC layers. In addition, to optimize the network training, the image input is processed through a robust strategy that subtracts the mean red-green-blue value of each pixel over the training set to improve training speed and accuracy. Furthermore, the authors have created a clouds dataset, called Cirrus Cumulus Stratus Nimbus (CCSN), which consists of 11 categories under meteorological standards.

In \cite{Deepmultimodal} authors propose an approach named deep multimodal fusion (DMF). In order to learn the visual features, CNN models have been applied to capture texture information. The extracted features, from deeper layers, have several eligible properties such as invariance and discrimination. Subsequently authors employ a weighted strategy to integrate visual and multimodal features. Finally, SVM algorithm to train the classification model is adopted.

In \cite{Deeptensor} a deep tensor fusion network is presented in order to hold spatial information of ground-based cloud images. It fuses cloud visual and multimodal features at the tensor level.

In \cite{HierarchicalMultimodal} author propose a approach, called Hierarchical Multimodal Fusion (HMF), which fuses deep multimodal and deep visual features in different levels. The architecture is composed of two subnetworks, visual subnetwork and multimodal subnetwork. The visual subnetwork is defined in order to extract deep visual features from ground-based cloud images employing Resnet50 \cite{he2016deep}. The multimodal subnetwork is used to learn features from a vector composed of six FC layers. Classification step through SVM is managed. 

In \cite{LBP2} author propose a classification method of sky-condition based on whole sky infrared cloud images, where the Local Binary Patterns operator (LBP) and the contrast of local cloud image texture (VAR signal) are combined to classify sky conditions. The correspondence relationship among traditional cloud classes and instrument-measured cloud classes is suggested. The approach analyzes the LBP spectra and VAR characteristics for five classes of clouds.

An automatic cloud classification algorithm is developed in \cite{Automaticcloud}, the approach uses a image-mask created by visually identifying image regions containing discriminative information. Furthermore the approach extracts a set of mainly statistical features describing the color as well as the texture of an image. Classification step adopts the k-nearest neighbour algorithm.

A modified texton-based classification approach that integrates both color and texture information to improve classification results is proposed in \cite{dev2015categorization}. Color channel is adopted to generate image descriptors and filter responses of images across all the categories aggregating them together. K-means clustering is applied on the concatenated filter responses, producing the different cluster centers. These clusters centers are the modified-textons and constitute the texton dictionary. The discriminative histogram model for each image category is generated by comparing the filter responses of the pixels with the generated textons in the dictionary.

An ensemble learning method and resource allocation scheme for cloud observation and classification is proposed in \cite{zhang2020ensemble}. Ensemble methods, like Bagging, AdaBoost and Snapshot are used as a base classifier to take the cross-semantic and structure features of cloud images.

\section{Materials and Methods}
\label{MM}
In this section we describe the proposed framework which includes two methodologies: deep neural networks \cite{liu2017survey} and voting learning \cite{peteiro2013survey}. The goal is to combine several deep neural networks with purpose to classify clouds images. Specifically, a set of competitive models are aligned and provide a range of confidential decisions useful for making choices during classification. The framework is composed of three blocks. A first, which performs preprocessing in terms of image resize. A second, which learns different deep neural networks, previously redesigned for the specific task. A third, which combines different potential indications, through voting rules, provided by deep neural networks for classification purpose. Finally, the framework runs a predetermined number of iterations in a supervised learning context.

\subsection{Image resize}
\label{resize}

One of the drawbacks of neural networks concerns the fixed dimension about the input layer with reference to the images to be processed (details about adopted neural networks can be found in table \ref{nets} at column 5). Size normalization, according to the input layer dimension, is essential because it is not possible to process different or large sized images for the network training and classification stages. This step does not alter the content of the image information in any way. 

\subsection{Network design and transfer learning}
\label{netmod}

The transfer learning has been selected as training strategy. The basic idea is to transfer the knowledge extracted from a source domain to a destination one, in our case clouds classification. Generally, a pretrained network is chosen as starting point in order to learn a new task. It turns out to be the most convenient and forthcoming solution to adopt the representational power of pretrained deep neural networks. Clearly, it is easy and fast to tune a network with transfer learning than training a new network from scratch with randomly initialized weights. For clouds recognition, deep learning architectures are selected based on their task compliance. The goal is to train networks on images by redesign their structures in the final layer according to different outgoing classes. Table \ref{nets} supports the description below about adopted neural models.

Alexnet \cite{krizhevsky2012imagenet} consists of 5 convolutional layers and 3 fully connected layers. It includes the non-saturating ReLU activation function, better then tanh and sigmoid during training phase.

Googlenet \cite{szegedy2015going} is composed of 22 deep layers. The network is inspired by LeNet \cite{lecun1989backpropagation} but implemented a novel element which is dubbed an inception module. This module is based on several very small convolutions in order to drastically reduce the number of parameters. The architecture reduced the number of parameters from 60 million (AlexNet) to 4 million. Furthermore, it includes batch normalization, image distortions and Root Mean Square Propagation algorithm.

Densenet201 \cite{huang2017densely} is a convolutional neural network with 201 deep layers. Unlike standard convolutional networks composed of $L$ layers with $L$ one-to-one connections between the current layers and the nexts, it contains $\frac{L(L+1)}{2}$ direct connections. Specifically, each layer adopts the feature-maps of all preceding layers and its own feature-maps into all subsequent layers as inputs.

Resnet18 and Resnet50 \cite{he2016deep} are inspired by pyramidal cells contained in the cerebral cortex. They use particular skip connections or shortcuts to jump over some layers. They are composed of 18 and 50 deep layers, which with the help of a technique known as skip connection has paved the way for residual networks.

Nasnetlarge \cite{zoph2018learning} is designed on a search space, called NASNet search space, which enables transferability. The model works by looking for the best convolutional layer, or cell, and subsequently replicating this layer in a stack, each with its own parameters to design a convolutional architecture. Also, a regularization technique, called ScheduledDropPath, that significantly improves generalization in the model is introduced.

\begin{table}[!ht]
\centering \caption{Description of adopted pretrained network.}
\footnotesize
\begin{tabular}{|l| l| l| l| l|}
\hline
  Network & Depth & Size (MB) & Parameters (Millions) & Input Size\\ \hline
  Densenet201 & 201 & 77 & 20 & 224 $\times$ 224\\ \hline
  Alexnet	& 8	& 227	& 61 & 227 $\times$ 227\\	
  \hline
  Googlenet	& 8	& 27 & 7 & 224 $\times$ 224\\
  \hline
  Resnet18 & 18 & 44 & 11.7 & 224 $\times$ 224\\	\hline
Resnet50 & 50 & 96 & 25.6 & 224 $\times$ 224\\
\hline
Nasnetlarge & * & 332 & 88.9 & 331 $\times$ 331\\
  \hline
\end{tabular}
\label{nets}
\end{table}

Deep neural networks have been adapted to the clouds classification problem. Originally, Imagenet dataset \cite{deng2009imagenet}, which includes one million images divided into 1000 classes, is adopted to perform the main training phase. Generally, a network elaborates an image and provides a prediction about a class it might belong to with an attached probability. Indeed, a network is structured to work on different layers. The first concerns the input image and requires 3 color channels. Next, convolutional layers, which work with the purpose to extract image features, are placed. The last learnable and the final classification layers are adopted to classify the input image. To make the pretrained network compliant to the classification of new images, the last two layers are replaced with new layers. Often, the last layer, with its learnable weights, is completely connected. It is removed and replaced by a new one completely connected with the outputs related to classes of new data (clouds types). Furthermore, the learning of the new layer, connected with the transferred layers, can be speeded up by increasing the learning rate factors. Optionally, the weights of the previous levels can be left unchanged by resetting the learning rate to zero. This modification avoids weights update during training and a consequent flattening of the execution time as it is not necessary to calculate the gradients of the relative layers. This improvement has a strong impact in the case of small datasets to avoid overfitting.

\subsection{Voting based learning}
\label{ense}

A voting based learning approach is adopted to manage the classification phase. In particular, among all possible strategies, we selected stacking. It works by training a single classifier and, subsequently, combines it with further classifiers. Unlike a standard approach, where weak or strong learners are adopted, we basically combined several equally powerful models that predict an outcome with a certain probability. Finally, we joined all the predictions for a classification result. The general model can be summarized by the following matrix

\begin{equation}
CN = \begin{bmatrix} 
    \beta_{1}i_{1} & \dots & \beta_{1}i_{k} \\
    \vdots & \ddots & \\
    \beta_{n}i_{1} &        & \beta_{n}i_{k} 
    \end{bmatrix}
\end{equation}

each $i_{k}$ represent an image to be classified, taken from the set $Imgs=\{i_{1},i_{2},\ldots,i_{k}\}$ with cardinality $k$, belonging to one of $x$ classes. Furthermore, each $\beta_{n}$ represent a deep neural network, taken from the set $C=\{\beta_{1},\beta_{2},\ldots,\beta_{n}\}$ with cardinality $n$, which provides a decision $d \in I \{1,\dots,x\}$, with reference to $i_k \in Imgs$ and $x$ membership classes. The set of decisions can be rearranged through the following matrix $D$

\begin{equation}
D = \begin{bmatrix} 
    d_{\beta_{1}i_{1}} & \dots & d_{\beta_{1}i_{k}} \\
    \vdots & \ddots & \\
    d_{\beta_{n}i_{1}} &        & d_{\beta_{n}i_{k}} 
    \end{bmatrix}
\end{equation}

it describes the result of deep neural networks combination and images of the matrix $CN$ in terms of position, such as $\beta_{n}i_{k}\to d_{\beta_{n}i_{k}}$. In addition, a score value $s \in S \{0,\dots,1\}$ is associated to each decision $d$ and provides the posterior probability $P(i|x)$ that an image $i$ could belong to class $x$. Finally, all the score values relating to the results of the possible combinations of matrix $CN$ are collected in the matrix $S$

\begin{equation}
S = \begin{bmatrix} 
    P(i_{1}|x)_{d_{\beta_{1}i_{1}}} & \dots & P(i_{k}|x)_{d_{\beta_{1}i_{k}}} \\
    \vdots & \ddots & \\
    P(i_{1}|x)_{d_{\beta_{n}i_{1}}} &        & P(i_{k}|x)_{d_{\beta_{n}i_{k}}} 
    \end{bmatrix}
\end{equation}

each element of posterior probability in the matrix $S$ refers to element of the matrix $CN$, such as $\beta_{n}i_{k}\to d_{\beta_{n}i_{k}} \to P(i_{k}|x)_{d_{\beta_{n}i_{k}}}$. 
Moving on, each column of the matrix $D$ is analyzed with statistical mode and stored in the vector $DM$

\begin{equation}
DM=\{dm_{d_{\beta_{1,\dots,n}i_{1}}},\dots, dm_{d_{\beta_{1,\dots,n}i_{k}}}\},\end{equation}

the generic value $dm$ contains the modal value of the class to which image $i$ could belong with the average probability score $ds$. In essence, this is the class to which an image could belong based on the votes given by different deep neural networks. In this regard, the concept of statistical mode is introduced. It can be defined as the value which is repeatedly occurred in a given set

\begin{equation}
mode=l+\left(\frac{f_1-f_0}{2f_1-f_0-f_2}\right)
\times h
\end{equation}

where $l$ is the lower limit of the modal class, $h$ is the size of the class interval, $f_1$ is the frequency of the modal class, $f_0$ is the frequency of the class which precedes the modal class and $f_2$ is the frequency of the class which successes the modal class. The columns of matrix $D$ are analyzed in order to obtain the values of the most frequent decisions. This step is performed in order to verify the highest voted classes from different deep neural networks, contained in the $CN$ set. Moreover, the aim of mode application is twofold. First, to extract the most frequent value. Second, to extract its occurrences in terms of indices. For each most frequent occurrence, modal value, the corresponding score from the matrix $S$ is extracted. To this end, $DS$ vector is built
 
\begin{equation}
DS=\{ds_{P(i_{1}|x)_{d_{\beta_{1,\dots,n}i_{1}}}}, \ldots, ds_{P(i_{k}|x)_{d_{\beta_{1,\dots,n}i_{k}}}} \},\end{equation}

where each element $ds$ contains the average decision scores with higher frequency, extracted through the mode, with reference to corresponding column of the matrix $D$.

\section{Experimental results}
\label{discussion}

This section describes the experimental phase. In order to train the neural models, with purpose to perform classification task in a supervised context, labeled data are need. Consequently, the issue to be addressed concerns the quantity of data sufficient to produce experimental results. The content of a large dataset, useful to training and testing, strongly affects the classification performance. Therefore, the discriminating factor about the effectiveness of neural models is the amount of data. Contextually, with purpose to produce compliant performance, the settings reported in recent cloud classification methods are adopted. 

\subsection{Datasets}
\label{dataset}

The proposed framework on a state-of-art datasets, containing ground-based clouds images, is tested. Datasets adopted are:
\begin{enumerate}
    \item Multimodal-Ground-based-Cloud-Database (MGCD) \cite{liu2020ground,liu2020multi}. It is collected in China and consists in cloud images captured by a sky camera with a fisheye lens under a variety of conditions and multimodal cloud information. It includes a total amount of 1720 cloud data. Images are divided into seven classes: cumulus, cirrus, altocumulus, clear sky, stratus, stratocumulus, cumulonimbus. The number of item of each class varies from 140 to 350, and the detailed numbers are listed in Table \ref{MGCdet}.

\item Singapore Whole sky IMaging CATegories Database  (SWIMCAT) dataset \cite{dev2015categorization}. It is composed of 784 sky/cloud patch images with 125 x 125 pixels captured using wide angle high-resolution sky imaging system, a calibrated ground-based WSI designed by \cite{WAHRSIS}. The dataset is splitted into five distinct categories: clear sky, patterned clouds, thick dark clouds, thick white clouds, and veil clouds. Details are present in table \ref{SWIdet}.

\item Cirrus Cumulus Stratus Nimbus (CCSN) dataset \cite{CloudNet}. It contains only 2,543 unique cloud images with 256 x 256 pixels in the JPEG format and contains 10 different forms in cloud observation. It is characterized by a large set of images, making it the largest of the available public cloud dataset. Details are shown in table \ref{CCSNdet}.

\end{enumerate}

\begin{table}[!ht]
\centering \caption{Details of MGCD dataset.}
\begin{tabular}{lllllllllll}
 \hline  \multicolumn{3}{|l}{Label}          & \multicolumn{3}{|l|}{Cloud Type} & \multicolumn{3}{l|}{Number of samples}  \\
 \hline \multicolumn{3}{|l}{1}  & \multicolumn{3}{|l|}{Cumulus} & \multicolumn{3}{l|}{160}  \\
\hline \multicolumn{3}{|l}{2}  & \multicolumn{3}{|l|}{Cirrus} & \multicolumn{3}{l|}{300}  \\
\hline \multicolumn{3}{|l}{3}  & \multicolumn{3}{|l|}{Altocumulus} & \multicolumn{3}{l|}{340}  \\
\hline \multicolumn{3}{|l}{4}  & \multicolumn{3}{|l|}{Clear sky} & \multicolumn{3}{l|}{350}  \\
\hline \multicolumn{3}{|l}{5}  & \multicolumn{3}{|l|}{Stratocumulus} & \multicolumn{3}{l|}{250}  \\
\hline \multicolumn{3}{|l}{6}  & \multicolumn{3}{|l|}{Stratus} & \multicolumn{3}{l|}{140}  \\
\hline \multicolumn{3}{|l}{7}  & \multicolumn{3}{|l|}{Cumulonimbus} & \multicolumn{3}{l|}{180}  \\
\hline

\end{tabular}
\label{MGCdet} 
\end{table}

\begin{table}[!ht]
\centering \caption{Details of SWIMCAT dataset.}
\begin{tabular}{lllllllllll}
 \hline  \multicolumn{3}{|l}{Label}          & \multicolumn{3}{|l|}{Cloud Type} & \multicolumn{3}{l|}{Number of samples}  \\
 \hline \multicolumn{3}{|l}{A}  & \multicolumn{3}{|l|}{Clear Sky} & \multicolumn{3}{l|}{224}  \\
 \hline \multicolumn{3}{|l}{B}  & \multicolumn{3}{|l|}{Patterned clouds} & \multicolumn{3}{l|}{89}  \\
 \hline \multicolumn{3}{|l}{C}  & \multicolumn{3}{|l|}{Thick dark clouds} & \multicolumn{3}{l|}{251}  \\
 \hline \multicolumn{3}{|l}{D}  & \multicolumn{3}{|l|}{Thick white clouds} & \multicolumn{3}{l|}{135}  \\
 \hline \multicolumn{3}{|l}{E}  & \multicolumn{3}{|l|}{Veil clouds} & \multicolumn{3}{l|}{85}  \\
\hline
\end{tabular}
\label{SWIdet} 
\end{table}

\begin{table}[!ht]
\centering \caption{Details of CCSN Dataset.}
\begin{tabular}{lllllllllll}
 \hline  \multicolumn{3}{|l}{Label}          & \multicolumn{3}{|l|}{Cloud Type} & \multicolumn{3}{l|}{Number of samples}  \\
 \hline \multicolumn{3}{|l}{Ci}  & \multicolumn{3}{|l|}{Cirrus} & \multicolumn{3}{l|}{139}  \\
\hline \multicolumn{3}{|l}{Cs}  & \multicolumn{3}{|l|}{Cirrostratus} & \multicolumn{3}{l|}{287}  \\
\hline \multicolumn{3}{|l}{Cc}  & \multicolumn{3}{|l|}{Cirrocumulus} & \multicolumn{3}{l|}{268}  \\
\hline \multicolumn{3}{|l}{Ac}  & \multicolumn{3}{|l|}{Altocumulus} & \multicolumn{3}{l|}{221}  \\
\hline \multicolumn{3}{|l}{As}  & \multicolumn{3}{|l|}{Altostratus} & \multicolumn{3}{l|}{188}  \\
\hline \multicolumn{3}{|l}{Cu}  & \multicolumn{3}{|l|}{Cumulus} & \multicolumn{3}{l|}{182}  \\
\hline \multicolumn{3}{|l}{Cb}  & \multicolumn{3}{|l|}{Cumulonimbus} & \multicolumn{3}{l|}{242}  \\
\hline \multicolumn{3}{|l}{Ns}  & \multicolumn{3}{|l|}{Nimbostratus} & \multicolumn{3}{l|}{274}  \\
\hline \multicolumn{3}{|l}{Sc}  & \multicolumn{3}{|l|}{Stratocumulus} & \multicolumn{3}{l|}{340}  \\
\hline \multicolumn{3}{|l}{St}  & \multicolumn{3}{|l|}{Strtus} & \multicolumn{3}{l|}{202}  \\
\hline \multicolumn{3}{|l}{Ct}  & \multicolumn{3}{|l|}{Contrails} & \multicolumn{3}{l|}{200}  \\
\hline

\end{tabular}
\label{CCSNdet} 
\end{table}

\subsection{Discussion}
\label{Discussion}
Table \ref{netadop} provides indications about the adopted neural models with respect to datasets. As can be seen, different combinations are provided due to the variable composition (total images, images per class, etc) of each datasets. In fact, two additional neural models, Resnet18 and Nasnetlarge, have been added for CCSN processing as it is the most complex to manage. Moreover, the framework consists in different modules written in Matlab language. Neural models were trained based on different parameters. Stochastic Gradient Descent (SGDM) with Momentum is adopted as solver of training process. Its main peculiarity concerns the oscillation along the steepest descent path towards the optimum. Adding a momentum term to the parameter update is one way to reduce this oscillation. Carrying on, MiniBatchSize value, the subset size of the training set adopted to evaluate the gradient of the loss function and update the weights, has been set to 10, optimal for the obtained results. About MaxEpochs, maximum number of epochs to use for training, the right compromise was reached with the value 6 optimizing execution time and performance. An iteration is a step performed by SGDM to minimize the loss function using MiniBatchSize. An epoch concerns the complete cycle of the training process on training set. InitialLearnRate has been set to 3e-4. If it is too low results in a high training time. Otherwise, if too high, the result may be suboptimal or training may diverge. The right compromise has also been found for the latter. To avoid discarding the same data every epoch, Shuffle parameter been set to every-epoch. Finally, ValidationFrequency, number of iterations between evaluations of validation metrics, has been set as the ratio between training set size and MiniBatchSize.

\begin{table}[!ht]
\caption{Deep neural networks adopted with respect to datasets.}
\centering
\begin{tabular}{|l|l|l|l|} \hline

\backslashbox{Networks}{Datasets} & MGCD & SWIMCAT & CCSN\\ 
\hline Densenet201 & \ding{52} & \ding{52} & \ding{52}\\ 
\hline
Alexnet	& \ding{52}	& \ding{52} & \ding{52} \\	
\hline
Googlenet	& \ding{52}	& \ding{52} & \ding{52}\\
\hline
Resnet18 &  &  & \ding{52}\\	
\hline
Resnet50 & \ding{52} & \ding{52} & \ding{52}\\
\hline
Nasnetlarge &  &  & \ding{52}\\
\hline
  \end{tabular}
\label{netadop}
\end{table}

The classification accuracy on MGCD dataset is presented in table \ref{compMGCD}. In order to produce a comparison with further methods, that work on the same ground-based cloud classification task, the settings described in \cite{liu2020multi} have been adopted. Looking at the results, several conclusions can be drawn. The proposed model, composed of different pretrained networks, produces promising recognition accuracy, giving a high representation of clouds images. The implemented voting based architecture compared with the competitors gets the better performance. 

\begin{table}[!ht]
\centering \caption{Experimental results on MGCD dataset.}
\begin{tabular}{lllllllllll}
 
                    \\
 \hline  \multicolumn{3}{|l|}{Method}          & \multicolumn{8}{l|}{Acc}   \\
\hline \multicolumn{3}{|l}{Our}  & \multicolumn{8}{|l|}{99.98}  \\
\hline \multicolumn{3}{|l}{MMFN \cite{liu2020multi}}  & \multicolumn{8}{|l|}{88.63}  \\
\hline \multicolumn{3}{|l}{DCAFs + MI \cite{liu2020multi}} & \multicolumn{8}{|l|}{82.97}  \\
\hline \multicolumn{3}{|l}{BOVW + MI  \cite{liu2020multi}} & \multicolumn{8}{|l|}{67.20}  \\
\hline \multicolumn{3}{|l}{PBOVW + MI  \cite{liu2020multi}} & \multicolumn{8}{|l|}{67.15}  \\
\hline \multicolumn{3}{|l}{LPB+ MI  \cite{liu2020multi}} & \multicolumn{8}{|l|}{50.53}  \\
\hline \multicolumn{3}{|l}{CLPB+ MI  \cite{liu2020multi}} & \multicolumn{8}{|l|}{69.68}  \\
\hline \multicolumn{3}{|l}{CloudNet + MI \cite{liu2020multi}} & \multicolumn{8}{|l|}{80.37}  \\
\hline \multicolumn{3}{|l}{BoVW \cite{Visualcategorization}}  & \multicolumn{8}{|l|}{66.15}  \\
\hline \multicolumn{3}{|l}{PBoVW \cite{Visualcategorization}}  & \multicolumn{8}{|l|}{66.13}  \\
\hline \multicolumn{3}{|l}{LBP \cite{LBP}}  & \multicolumn{8}{|l|}{55.20}  \\
\hline \multicolumn{3}{|l}{CLBP \cite{CLBP}}  & \multicolumn{8}{|l|}{69.18}  \\
\hline \multicolumn{3}{|l}{VGG-16 \cite{simonyan2014very}}  & \multicolumn{8}{|l|}{77.95}  \\
\hline \multicolumn{3}{|l}{DCAFs \cite{DCAFS}} & \multicolumn{8}{|l|}{82.67}  \\
\hline \multicolumn{3}{|l}{CloudNet \cite{CloudNet}} & \multicolumn{8}{|l|}{79.92}  \\

\hline \multicolumn{3}{|l}{DMF  \cite{Deepmultimodal}} & \multicolumn{8}{|l|}{79.05}  \\
\hline \multicolumn{3}{|l}{DTFN  \cite{Deeptensor}} & \multicolumn{8}{|l|}{86.48}  \\
\hline \multicolumn{3}{|l}{HMF  \cite{HierarchicalMultimodal}} & \multicolumn{8}{|l|}{87.90}  \\

\hline
\end{tabular}
\label{compMGCD} 
\end{table}

The classification performance on SWIMCAT dataset are summarized in table \ref{compSWIMCAT}. In this phase, the settings present in \cite{DCAFS} are adopted. In particular, a cross validation on 2,3,4,5 folds was performed first. Subsequently, 40 images per class for training and 45 ones for testing have been selected randomly. The average accuracy of 50 random runs is reported. In the training phase, the neural models that did not contribute to improve both the performance and the execution time were discarded, as can be seen in table \ref{netadop}. The results highlight that combining different multiple classification predictions is useful to capture more spatial and local layout information of clouds with purpose to outperform the compared methods. Furthermore, it is important to underline that the proposed approach is even better than neural models, such as VGG-16 \cite{simonyan2014very} and CloudNet \cite{CloudNet}, in which a single classification confidence value is provided compared to a multiple voting based mechanism.

\begin{table}[!ht]
\caption{Experimental results on SWIMCAT dataset.}
\centering
\begin{tabular}{|l|l|l|l|l|l|} \hline

\backslashbox{Methods}{Folds} & 2 & 3 & 4 & 5 & 40/45\\ \hline
Our & 99.36 & 99.49 & 99.49 & 99.75 & 99.91\\ 
\hline
LPB \cite{LBP2} & 85.26 & 81.60 & 83.51 & 85.03 & 93.47\\ 
\hline
Heinle Feature \cite{Automaticcloud} & 90.26 & 91.89& 92.91 & 93.43 & 93.09\\ 
\hline
Text-based method  \cite{dev2015categorization} & - & - & - & - & 95.00\\ 
\hline
DCAF  \cite{DCAFS} & 98.72 & 98.46 & 98.97& 98.84 & 99.56\\ 
\hline
\end{tabular}
\label{compSWIMCAT}
\end{table}

Table \ref{compCCSN} shows the performance on CCSN dataset. In order to compare results with further methods that work on same task, the settings described in \cite{zhang2020ensemble} have been adopted. Also in this case, the combination of neural models lead to better performance. As shown in table \ref{netadop}, for this experimental phase we have stacked all analyzed deep neural network. Once again, the results demonstrate that the multiple base learners may lead to better performance according to the combination with the different number of base learners in the stacking.

\begin{table}[!ht]
\centering \caption{Experimental results on CCSN dataset.}
\begin{tabular}{|l| l|}
 \hline Method           & Acc   \\
\hline Our  & 95.08  \\
\hline Cloudnet \cite{CloudNet} & \ 90.00  \\
\hline \cite{zhang2020ensemble} & 80.00\\
\hline MMI \cite{MultimodalGroundBased}  & 75.42\\
\hline M\textunderscore DF \cite{MultimodalGroundBased} & 78.21\\
\hline M\textunderscore JFCNN\cite{MultimodalGroundBased}  & 84.55\\
\hline V\textunderscore DF \cite{MultimodalGroundBased}  & 85.10\\
\hline V\textunderscore JFCNN \cite{MultimodalGroundBased}  & 86.79\\
\hline V\textunderscore DF + MMI \cite{MultimodalGroundBased}  & 86.33 \\
\hline V\textunderscore JFCNN + MMI \cite{MultimodalGroundBased}  & 89.40\\
\hline V\textunderscore DF + M\textunderscore DF \cite{MultimodalGroundBased}  & 90.21 \\
\hline J\textunderscore JFCNN \cite{MultimodalGroundBased} & 78.82\\
\hline JFCNN \cite{MultimodalGroundBased}  & 93.37 \\
\hline
\end{tabular}
\label{compCCSN}
\end{table}

The presented satisfactory results are attributable to many relevant aspects. The first regards the features extracted through convolutional layers of the deep neural network. They provide a good image representation, although are completely abstract and devoid of real meaning. Second regards the framework capability to provide multiple representation models, that lead to a significant improvement in performance. Another issue concerns the image size normalization, tackled to many methods in the field. It is performed before the features extraction, to avoid performance degradation. Again, we can look at the robustness with respect to the underrepresented classes in the datasets. In fact, the framework does not fail even though the samples are not sufficient for a class representation in specific cases. The latter appears to be an open problem in the literature, as ad hoc classifiers are often designed for unbalanced classification different from the standard ones that produce untrue results. Contrary, a weak point concerns the computational aspect. First, the time required for training pretrained model is high but less than a model created from scratch. Second, the classification step, that provides multiple choices in decision making at each iteration, requires a lot of effort. The latter works for the purpose of choosing which classifiers are suitable for specific clouds images included in the test set. Finally, we have shown that although the framework is more expensive from a computational point of view and it produces better results than a single classifier.

\section{Conclusions and Future Works}
\label{conc}

The challenge in ground-based cloud recognition is specifically interesting and, not only, for its multiple aspects and variety of data. The complexity of the task is linked to several factors such as the type of clouds and the visual patterns contained in them. In support, convolutional neural networks lend a big hand to understand the meaning of images with the consequent goal of their classification. In this regard, we proposed a framework that combines convolutional neural networks, adapted to the cloud recognition task through a transfer learning approach, using voting rules. The results produced certainly strengthen the theoretical thesis. A multiple model, based on several deep neural networks, compared to a single one is a powerful factor. Through a large experimental phase, it has been shown how the proposed approach is competitive, and in some cases better, compared to the more advanced methods. Although pretrained models have been adopted, the main weakness concerns the computational complexity of learning phase that requires a long time, sensitive to the growth of the data. Future work will certainly concern the study and analysis of still unexplored convolutional neural networks for this type of problem and the application of the proposed framework to further datasets with the aim of taking a step forward in cloud recognition.

\section*{Acknowledgements}
Our thanking is for Alfredo Petrosino. He followed us during the first steps towards the Computer Science, through a whirlwind of goals, ideas and, especially, love and passion for the work. We will be forever grateful great master.



\bibliographystyle{elsarticle-harv}
\bibliography{bibliography}





\end{document}